\documentstyle[twoside,fleqn,jhuwkshp,psfig]{article}
\newcommand{\etal}{{\it et al.} }
\newcommand{\asca}{{\it ASCA} }

\begin{document}

\title{Fe K EMISSION LINES IN LOW-LUMINOSITY AGNS}
 
\author{Y. Terashima,
\address{The Institue of Space and Astronautical Science}
\address{Astronomy Department, University of Maryland, College Park, MD 20742, USA} 
N. Iyomoto$^{\rm a}$,
L.~C. Ho, \address{The Observatories of the Carnegie Institution of Washington}
A.~F. Ptak \address{Department of Physics and Astronomy,
Johns Hopkins University, 3400 N. Charles Street, Baltimore, MD 21218, USA.}
}

\begin{abstract}
We present the results of a systematic analysis of Fe K emission lines
in low-luminosity AGNs (LLAGNs) observed with {\it ASCA}. We used a
sample of LLAGNs with small intrinsic absorption to compare with
luminous AGNs and accretion models. Some objects show an Fe line at
6.4 keV or 6.7 keV, and the rest of the objects show no significant Fe
K emission line.  We made a composite spectrum of sources without
detected Fe emission lines to search for weak lines and obtained an
upper limit on the equivalent width (EW) of a narrow line at 6.4 keV
of 145 eV, which is marginally consistent with EWs seen in Seyfert 1s. 
Even in those objects with detected Fe lines, a skewed broad
component, typically observed in Seyfert 1s, is not seen. The weakness
of the Fe line is consistent with the absence of an optically thick
disk in the vicinity of a central black hole. Alternatively, it might be
due to the ionization of a standard disk. The lack of big blue bump
favors the former.

\end{abstract}

\maketitle

\section{Introduction}

Optical spectroscopic surveys have shown that more than one third of
nearby galaxies harbor low-luminosity AGNs (LLAGNs;
\cite{Ho1997}). Their bolometric luminosities ($L_{\rm bol}<10^{42}$
ergs s$^{-1}$) and the black hole masses measured dynamically indicate
that LLAGNs are radiating at an extremely low Eddington ratio,
typically $L/L_{\rm Edd} \approx 10^{-3}-10^{-6}$. Models of
low-radiative efficiency accretion flows such as advection-dominated
accretion flows (ADAFs) have been developed and applied to LLAGNs
(e.g., \cite{Quat2001}).

Fe K emission lines are a powerful diagnostic tool to probe the inner
parts of accretion disks. In the ADAF scenario, the inner parts of the
disk are almost fully ionized, and no Fe K emission line is expected. On
the other hand, if an optically thick standard disk is present, a
fluorescent line is produced. In this paper, we summarize \asca
results on Fe K lines and give constraints on accretion disks 
in LLAGNs.

\section{The Sample}

We systematically analyzed 52 {\it ASCA} observations of 21 LINERs
(low-lonization nuclear emission-line regions) and 17 low-luminosity
Seyferts (\cite{Tera2002a}). The sample is constructed primarily based
on the classification of the Palomar survey (\cite{Ho1997}), with 
several other objects added. We made a subsample of LLAGNs from this
sample by choosing objects whose hard X-ray emission is dominated
by the nucleus, by using X-ray variability, spectra, images, and X-ray
to optical emission ratios to judge whether the AGN contribution dominates
(\cite{Tera2002b}). Highly absorbed objects are
excluded since they are not suitable for studying accretion disks
because of the complicated Fe K emission expected from sources other sources.
The final sample consists of 14 observations of 13 LLAGNs.  The basic data of 
the galaxies in our sample are shown in Table 1.

%\begin{figure*}[t] % fig.2
%\vspace{10pt}
%{\psfig{file=jhuwkshp_fig.ps,width=6.0in,height=3.5in}}
%\caption{Large figure spreading across two columns.
%}\label{fig:largefig}
%\end{figure*}

\begin{table*}[tb]
\caption{\centerline{The Sample} }
\begin{center}
\begin{tabular}{lcccc}
\hline \hline
Name    & Class         & Distance	& $L_{\rm X}$ $^a$       & $L_{\rm X}/L_{\rm Edd}$\\
%        &              & (Mpc)	 & $10^{40}$ ergs s$^{-1}$       & \\
\hline
NGC 315         & L1.9  & 65.8	& 49    & ...\\
NGC 1097        & L/S1.5& 14.5	& 5.2   & $4.6\times10^{-6}$\\
NGC 3031        & S1.5  & 1.4	& 0.43  & $2.0\times10^{-7}$\\
NGC 3147        & S2    & 40.9	& 34    & $3.4\times10^{-6}$\\
NGC 3998        & L1.9  & 21.6	& 46    & $6.5\times10^{-6}$\\
NGC 4203        & L1.9  & 9.7	& 2.3   & $9.3\times10^{-5}$\\
%NGC 4258 (93) 	& S1.9  & 6.8	& 6.8   & $1.3\times10^{-5}$\\
%NGC 4258 (96) 	&       & 6.8	& 17    & $3.2\times10^{-5}$\\
NGC 4261        & L2    & 35.1	& 15    & $2.3\times10^{-6}$\\
NGC 4450        & L1.9  & 16.8	& 2.2   & $1.8\times10^{-6}$\\
NGC 4579$^b$ 	& S/L1.9& 16.8	& 14    & $2.3\times10^{-6}$\\
NGC 4579$^c$ 	&       & 	& 20    & $3.2\times10^{-6}$\\
NGC 4594        & L2    & 20.0	& 14    & $1.0\times10^{-6}$\\
NGC 4639        & S1.0  & 16.8	& 3.6   & ...\\
NGC 5005        & L1.9  & 21.3	& 3.8   & ...\\
NGC 5033        & S1.5  & 18.7	& 23    & $6.1\times10^{-5}$\\
\hline
\end{tabular}

$a$: luminosity in 2--10 keV band in units of $10^{40}$ ergs s$^{-1}$; $b$: 
observed in 1995; $c$: observed in 1998 \end{center}
\end{table*}

%MBH - sigma ralation: NGC 4203, 4579, 5033

\section{Results}

The X-ray spectra of LLAGNs are well fitted by a canonical model that
consists of a power law (photon index $\Gamma \approx1.8$) and a
Raymond-Smith thermal plasma with $kT\approx0.6$ keV.  The thermal
component was not required in five objects (NGC 3147, 3998, 
4203, 4639, and 5033), but the other objects in the present
sample were well described by the canonical model. Fe K emission is
detected in seven out of 13 objects. The line center energy, line
width, and equivalent width (EW) are summarized in Table 2. For
objects without significant Fe K emission, the line center energy was
fixed at 6.4 keV in the source rest frame and upper limits on EW were
calculated. The line center energies in three objects (NGC 3147, 3998,
and 5033) are 6.4 keV, which is expected from cold (or low-ionization
state) Fe, while two objects (NGC 4261 and 4639) have higher energies
($\sim$ 6.7 keV), consistent with He-like Fe. Although the center
energy of the Fe line in NGC 3031 (M81) is consistent with both cold
and highly ionized Fe, other observations of NGC 3031 indicate a higher
line centroid energy (\cite{Ishi1996}). NGC 4579 was
observed more than twice. The line center energy in this 
object decreased from 6.7 keV to 6.4 keV, accompanied by an increase of the
continuum luminosity by 35\%. We note that the emission lines at 6.7
keV most likely originate from an AGN, particularly in the case of NGC
3031 (\cite{Ishi1996}) and NGC 4579 (\cite{Tera1998,Tera2000}),
and a thermal origin is unlikely because the continuum of NGC 3031 has 
a power-law shape and the Fe line in NGC 4579 is variable.

Thus, Fe line energies in LLAGNs shows some diversity compared to luminous
Seyferts, in which the line is generally peaked at 6.4 keV. No broad, 
asymmetric Fe lines such as those seen in classical Seyfert 1s are detected in 
LLAGNs (e.g., \cite{Tera1998,Tera1999}), although the photon statistics are 
limited.

The upper limits on EW are not very tight for the objects without
significant Fe lines. In order to obtain tighter constraints, we made
a composite spectrum. We coadded the spectra of six objects (NGC 315,
1097, 4203, 4450, 4594, and 4639) from which Fe lines are not detected
or only marginally detected. All of these have very similar continuum
slopes.  NGC 5005 was not used because its spectral slope is flatter
than others. The composite spectrum, shown in Figure 1, is well fitted by the 
canonical model and no Fe line is required. The upper limit to the EW of a 
narrow Fe line at 6.4 keV is $38^{+107}_{-38}$
eV. This EW is smaller than or marginally consistent with that for a
core in the disk-line profile in luminous Seyfert 1s
(\cite{Nand1997}). The constraint on the EW of a broader component is 
weaker.  For example, the EW of a Gaussian line with $\sigma=0.2$ keV is
$74^{+137}_{-74}$ eV. This upper limit lies on the lowest end of the
distribution of Fe EWs in luminous Seyfert 1s.

The continuum of the composite spectrum is well described by a power
law with a photon index of $1.64\pm0.03$; a thermal Bremsstrahlung
model is not adequate. This reconfirms that the X-ray emission in the
present sample is dominated by an AGN and that the contribution from
discrete X-ray source such as low-mass X-ray binaries in the host galaxy is
small.  Therefore, the weakness of an Fe emission line is not due to
dilution by sources other than an AGN; instead, it appears to be a 
property intrinsic to some LLAGNs.

%The smaller EW suggest the solid angle of the cold matter seen from
%the X-ray source is smaller in these LLAGNs.

\begin{table*}[t]
\caption{\centerline{Results of Spectral Fits} }
\begin{center}
\begin{tabular}{lccccc}
\hline \hline
Name    & $N_{\rm H}$	& Photon Index	& Energy$^a$    & Width & EW \\
        & ($10^{20}$ cm$^{-2}$)	&		& (keV)         & (keV) & (eV)\\
\hline
NGC 315  & $0.49^{+0.41}_{-0.33}$ & $1.73^{+0.28}_{-0.25}$ & 6.4 & 0 & $<$380\\
NGC 1097 & $0.13^{+0.10}_{-0.07}$ & $1.67^{+0.09}_{-0.10}$& 6.4 & 0 & $<$160\\
NGC 3031 & 0($<0.053$)           & $1.796^{+0.027}_{-0.028}$ & $6.59^{+0.22}_{-0.13}$ & 0    & $106^{+59}_{-56}$\\
NGC 3147 & $0.062^{+0.05}_{-0.024}$ & $1.82^{+0.10}_{-0.09}$ & $6.49\pm0.09$ & 0     & $490^{+220}_{-230}$\\
NGC 3998 & $0.082\pm0.012$       & $1.90^{+0.03}_{-0.04}$ & $6.41^{+0.12}_{-0.19}$ &0     & $85^{+81}_{-71}$\\
NGC 4203 & 0.022($<$0.053)       & $1.78^{+0.07}_{-0.08}$ & 6.4 & 0 & $<$310\\
%NGC 4258 (93) & $6.66^{+0.20}_{-0.07}$ &0    & $108^{+73}_{-64}$\\
%NGC 4258 (96)       & $6.31^{+0.09}_{-0.10}$ &0     & $54^{+25}_{-27}$\\
NGC 4261 & 0.17($<$0.39)         & $1.30^{+0.07}_{-0.06}$ & $6.85^{+0.08}_{-0.15}$ & 0    & $550^{+300}_{-310}$\\
NGC 4450 & 0($<$0.082)           & $1.75^{+0.18}_{-0.17}$ & 6.4 & 0 & $<$1200\\
NGC 4579$^b$ & $0.04\pm0.03$     & $1.72\pm0.05$ & $6.73^{+0.13}_{-0.12}$ & $0.17^{+0.11}_{-0.12}$       & $490^{+180}_{-190}$\\
NGC 4579$^c$ & 0.04($<0.13$)     & $1.81\pm0.06$ & $6.39\pm0.09$  & 0 ($<$0.16)   &$250^{+105}_{-95}$ \\
NGC 4594 & $0.73\pm0.29$         & $1.89\pm0.16$ & 6.4 & 0     & $<$150\\
NGC 4639 & $0.069^{+0.041}_{-0.038}$ & $1.66\pm0.10$     & $6.67^{+0.16}_{-0.23}$ &0     & $520^{+320}_{-300}$\\
NGC 5005 & 0.10($<$0.86)         & $0.97\pm0.37$  & 6.4 & 0     & $<$810\\
NGC 5033 & $0.087\pm 0.017$      & $1.72\pm 0.04$ & $6.43^{+0.13}_{-0.08}$ & 0.08 ($<$0.23)       & $306^{+116}_{-119}$\\
\hline
\end{tabular}
\end{center}
Errors are at the 90\% confidence for one parameter of interest. The
data without errors denote fixed parameter. $a$: The line center
energy is in the source rest frame; $b$: observed in 1995; $c$:
observed in 1998.
\end{table*}

\begin{figure*}[t]
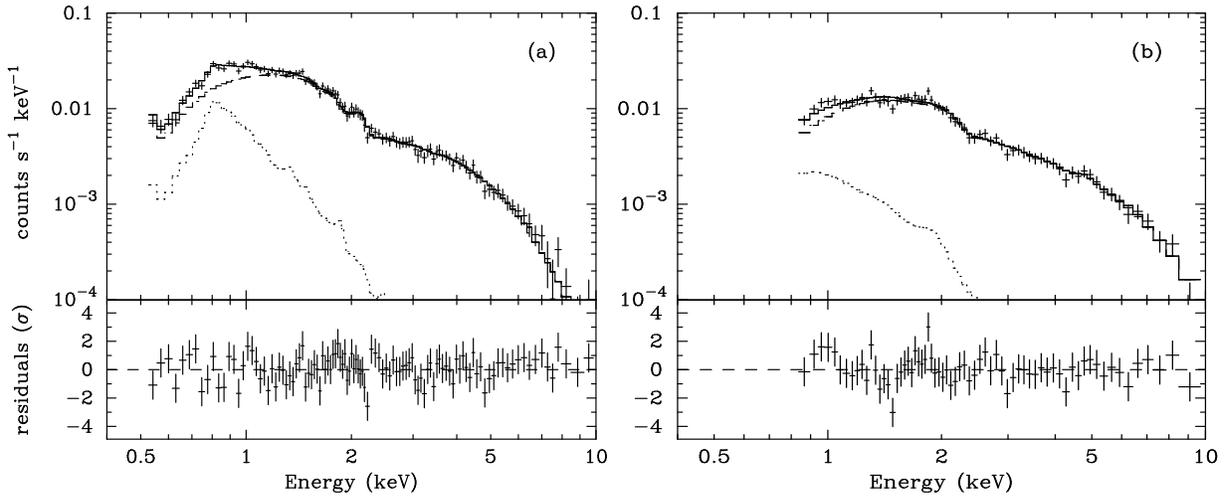
 % fig.1
\vspace{10pt}
\centerline{\psfig{file=terashima_y_fig1a.ps,angle=-90,width=8.0cm,height=6.5cm}
\psfig{file=terashima_y_fig1b.ps,angle=-90,width=8.0cm,height=6.5cm}
}
\caption{Composite (a) SIS and (b) GIS spectrum of six LLAGNs. Although spectral
fits were done simultaneously, the figures are shown separately for clarity.
}\label{fig:smallfig}
\end{figure*}

\section{Discussion}

  We have shown that Fe emission lines are seen in some LLAGNs but that they are 
very weak in others. Since we targeted objects whose hard X-ray emission is 
most likely dominated by the nucleus, the observed properties of the X-ray 
spectra are attributed to the nature of LLAGNs. The observed properties 
(continuum shape and Fe line parameters) show no clear correlation with $L_{\rm
X}$, black hole mass, and $L_{\rm X}/L_{\rm Edd}$. The continuum slope of
LLAGNs are in the range $\Gamma$ = 1.6--1.9, which is quite similar to those seen
in luminous AGNs. These photon indices are also in agreement with the
predictions of low-radiative efficiency accretion models, given the 
accretion rates of the present sample ($L_{\rm X}/L_{\rm Edd} \approx 
10^{-6}$; see Table 1) (e.g., \cite{Ball2001}).

%the continuum shape in the hard X-ray band provides no
%strong constraint on the structure of the accretion disks.

  The small EW of the Fe line in the composite spectrum is in
accordance with the absence of an inner optically thick
disk. Alternatively, the inner part of such a disk may be highly
ionized.

  If an optically thick standard disk is present, a broad Fe K line is
expected through reprocessing by the disk.  Depending on the accretion rate, 
the surface of the disk could be ionized by irradiation (e.g.,
\cite{Naya2000}). When the disk surface is almost completely ionized
to the Thomson depth, an Fe K line is no longer expected.  An Fe K
line can also be suppressed if the ionization state of Fe is
intermediate due to resonant trapping and Auger destruction
(\cite{Ross1993}). An emission line without a broad component can be
produced if the inner part of the disk is highly ionized.  Thus, the
observed Fe lines in LLAGNs (centered at 6.4 keV or 6.7 keV, or absent) might 
be understood in terms of the ionization state
of the disk. There are, however, serious problems in this scenario.
LLAGNs have very low luminosities and very low Eddington ratios 
(Table 1). It seems difficult to achieve high ionization states under 
these conditions.

The detected Fe lines that lack a very broad component are consistent with
a scenario in which an inner optically thick disk is absent. If the
inner part of the accretion disk becomes a hot ADAF, the Fe atoms should be 
completely ionized and no Fe emission line is expected.  In this case,
we might observe a fluorescent Fe line only from the outer optically thick
regions of the disk, or from regions exterior to the disk (e.g., an obscuring 
torus presumed in the unification scheme of Seyferts). The limits on the EW of 
a narrow line at 6.4 keV in the composite spectrum indicate that cold matter
(optically thick disk and torus) subtends only a small solid angle as viewed
from the X-ray source; this is consistent with the expectation.  An Fe
line at 6.4 keV is seen in four objects, and their relatively narrow
width is also in agreement with this scenario. An additional support
for the absence of an inner optically thick disk is the lack of big
blue bump (BBB; \cite{Ho1999,Quat1999}). Note, however, that the inner
temperature of an optically thick disk depends on the black hole mass
and mass accretion rate. A definitive evaluation of the strength of the 
BBB will require accurate black hole mass measurements 
(\cite{Quat2001}). Therefore, the combination of Fe K lines, spectral
energy distributions, and black hole mass measurements provides a
powerful diagnostic tool of accretion disks.  Although the absence of
Fe lines and Fe lines at 6.4 keV can be understood in a scenario invoking an 
ADAF plus an outer standard disk, the Fe lines at 6.7 keV are still
puzzling. Future observations of Fe lines with high signal-to-noise
ratio will be able to constrain the location of Fe-emitting region and
the structure of accretion disks in LLAGNs.

%do not change this

%do not change this
\normalsize

\section*{ACKNOWLEDGEMENTS}
 
Y.~T. and N.~I. are supported by the Japan Society for the Promotion of
Science Postdoctoral Fellowships for Young Scientists.  

\end{document}